# Tuning the torque-speed characteristics of bacterial flagellar motor to enhance the swimming speed

Praneet Prakash[a], Amith Z. Abdulla[b], Varsha Singh[c], Manoj Varma[a,d,*]

[*]mvarma@iisc.ac.in

[a]Centre for Nanoscience and Engineering, Indian Institute of Science, Bangalore, India; [b]Department of Physics, Indian Institute of Science, Bangalore, India; [c]Molecular Reproduction, Development and Genetics, Indian Institute of Science, Bangalore, India; [d]Robert Bosch Centre for Cyber Physical Systems, Indian Institute of Science, Bangalore, India.

**Abstract**

In a classic paper, Edward Purcell analysed the dynamics of flagellated bacterial swimmers and derived a geometrical relationship which optimizes the propulsion efficiency. Experimental measurements for wild-type bacterial species *E. coli* have revealed that they closely satisfy this geometric optimality. However, the dependence of the flagellar motor speed on the load and more generally the role of the torque-speed characteristics of the flagellar motor is not considered in Purcell's original analysis. Here we derive a tuned condition representing a match between the flagella geometry and the torque-speed characteristics of the flagellar motor to maximize the bacterial swimming speed for a given load. This condition is independent of the geometric optimality condition derived by Purcell and interestingly this condition is not satisfied by wild-type *E. coli* which swim 2-3 times slower than the maximum possible speed given the amount of available motor torque. Our analysis also reveals the existence of an anomalous propulsion regime, where the swim-speed increases with increasing load (drag). Finally, we present experimental data which supports our analysis.

**Manuscript**

In this Letter, we revisit the problem of propulsion efficiency of flagellated bacterial swimmers such as *E. coli* [1,2]. Flagellated bacteria propel themselves by generating torque using the bacterial flagellar motor (BFM) which rotates the helically shaped flagella [3]. The chirality of the helical shape results in coupling of the rotational and translational degrees of freedom resulting in propulsion of the cell body. The dynamics of a bacterium propelled by helical flagella (Fig. 1(a)) can be expressed by the propulsion matrix [4] which relates the force and torque on bacterial body and flagella with the speed and angular velocity. Force and torque on a bacterial body is given by:

$$\begin{pmatrix} F \\ \tau \end{pmatrix} = \begin{pmatrix} A_o & 0 \\ 0 & D_o \end{pmatrix} \begin{pmatrix} V \\ \Omega \end{pmatrix} \qquad (1)$$

$A_o$, $D_o$ are the translation and rotational drag coefficients and $V, \Omega$ are the translation and angular speed of bacterial body. To enforce the zero net force and torque condition, the propulsion matrix of the flagella satisfies:

$$\begin{pmatrix} -F \\ -\tau \end{pmatrix} = \begin{pmatrix} A & B \\ B & D \end{pmatrix} \begin{pmatrix} V \\ -\omega \end{pmatrix} \qquad (2)$$

Here, A, D are the translational and rotational drag coefficients and V, ω are the translation and angular speed of the flagella. The constant, B, couples the rotational motion of helical flagella to the translation motion of bacterium.

Purcell in his landmark paper on helical swimming considered the optimal geometry of the flagella which will maximize the propulsion efficiency η, defined as $\eta = \frac{A_0 V^2}{\tau \Omega_m}$ where $\Omega_m$ is the rotational speed of the flagellar motor, for a given size of cell body. Purcell showed that η is maximised when the translational drag coefficient of the flagella is matched to that of the cell body, i.e. $A = A_0$. The propulsion matrix elements in Eq. (1) and (2) have been explicitly measured for the bacterial species *E. coli* and A and $A_0$ are found to be quite close to each other ($1.48 \times 10^{-8}$ N.s.m$^{-1}$ and $1.4 \times 10^{-8}$ N.s.m$^{-1}$ respectively) [5] in agreement with Purcell's result. Equations (1) and (2) can be solved to obtain the swim-speed V and the torque τ produced by the flagellar motor in terms of the flagellar motor speed $\Omega_m = \Omega + \omega$ (see Sec. 1 of the Supplemental Material (SM)) for the detailed derivation).

$$V = \left[\frac{BD_o}{(A + A_o)(D + D_o)}\right] \Omega_m \quad (3)$$

$$\tau = \left[\frac{DD_o}{D + D_o}\right] \Omega_m \quad (4)$$

Under the assumption of $D \ll D_0$ which is satisfied by E. coli ($D = 7 \times 10^{-22}$ N.s.m and $D_0 = 6D = 4.2 \times 10^{-21}$ N.s.m) these equations simplify to $V = \frac{B}{A+A_0}\Omega_m$ and $\tau = D\Omega_m$. Under the optimum propulsion efficiency condition swimming speed becomes $V = \frac{B}{2A}\Omega_m$. Using $B = 7.9 \times 10^{-16}$ N.s and $\Omega_m = 850$ rad.s$^{-1}$ (~135 Hz), we get $V = 22.5$ μm/s, which is remarkably close to experimental observation. However, if we rewrite the equation of swim-speed in terms of torque as $V = \frac{B}{A+A_0}\frac{\tau}{D}$ and ask the question what is the maximum possible swim speed, we see that it depends on the maximum torque which can be produced by the BFM which is in the range of 1200 – 1500 pN-nm [6]. Using this value of maximum torque, we can estimate the maximum possible swimming speed, $V_{max}$, to be around 50 μm/s which is 2.5 times the observed swimming speed of E coli. This calculation implies that free-swimming bacteria do not tap the full torque available to them from the BFM and which then naturally leads to the questions namely: a) under what conditions can the bacteria tap the maximum torque available from the BFM to increase their swimming speeds and b) if this speed enhancement comes at the cost of sacrificing propulsion efficiency. We derive a simple relationship between the rotational drag coefficient of the flagella and shape of the torque-speed characteristic of the BFM which determines whether a bacterium can tap the maximum available torque. This relationship is independent of the optimum propulsion efficiency criterion, $A = A_0$. Therefore, the enhancement in speed arising from tapping the maximum torque from the BFM need not come at the cost of lowering the propulsion efficiency. These two conditions can be simultaneously satisfied. As we have already seen in the calculations above, wild-type bacteria such as *E. coli* and a common bio-film forming bacteria species *Psuedomonas aeruginosa* (as we show later in the article) do not tap the maximum torque available. In other words, these species, and perhaps other species as well, have evolved to swim at two to three times lower speeds than what would be possible if they used all the torque available from the BFM [7]. The reason for this is not clear at present. In general, we derive the conditions to achieve the maximum swimming speed for a flagellated bacterium with a specified load. The load could simply be the translational and rotational drag associated with swimming through the fluid or additional cargo attached to the bacterial cell body as in the case of recent examples of bacterial bio-hybrid robots [8–11]. Therefore, this work may also be useful for optimal design of future bio-hybrid robots. Our theoretical analysis additionally reveals an interesting 'anomalous swimming regime', where contrary to normal behaviour, swim speed increases with increasing load on the bacterial cell body. Lastly, we present experimental measurements of the speed *vs.* load curve for *Psuedomonas aeruginosa* to support our analysis. Demonstration of speed enhancement via utilizing the maximum BFM torque in real bacteria is currently beyond our capability as the relevant genetic circuits need to be identified and engineered to tune the torque-speed characteristic.

When a bacterium is loaded with some cargo on its cell body as depicted in Fig. 1(a), the drag coefficients $A_0$ and $D_0$ can be written as:

$$A_0 = \alpha A_0^{cell-body} + A_0^{cargo} \text{ and } D_0 = \alpha D_0^{cell-body} + D_0^{cargo} \qquad (5)$$

where, $A_0^{cargo}$ and $D_0^{cargo}$ for a spherical cargo of radius r is given by $6\pi\eta r$ and $8\pi\eta r^3$ respectively, and $\eta$ is the viscosity of the fluid medium. The factor $\alpha$ is to account for the partial occlusion of the cell body by the cargo [12,13]. In our simulations, we take $\alpha$ to be 0.3, although our conclusions are valid for any $\alpha$.

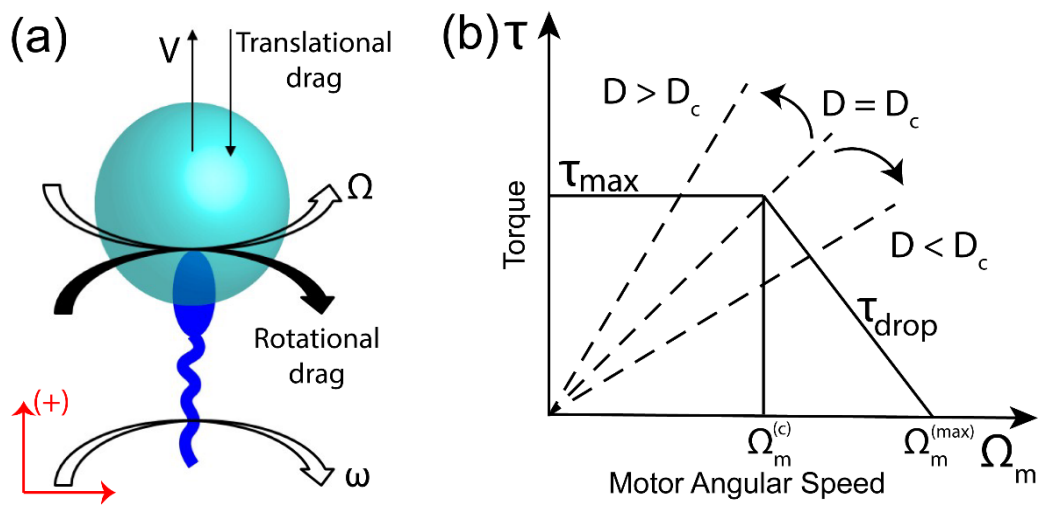

FIG. 1. Propulsion by the rotation of helical flagella. (a) Schematic depicting oil droplet as 'cargo' attached to the bacterial body. (b) Typical torque speed response curve of bacterial motor.

In order to estimate the swimming speed at a given load, $D_0$, we must determine the motor speed ($\Omega_m$) at the corresponding load from the torque-speed characteristic of the BFM. A typical torque-speed (TS) characteristic of a bacterial motor is shown in Fig. 1(b). The TS characteristics of the BFM is characterized by a "flat-torque" region of constant torque up to a knee frequency $\Omega_m^{(c)}$, beyond which the torque drops to zero approximately linearly till zero torque [6]. The motor speed for a given load is specified by the intersection of load-line $\tau = DD_0/(D + D_0)\Omega_m$ (Eq. 4) and the TS. As we see from Eq. 4, the slope of the load-line (Fig. 1(b)) keeps increasing for higher loads matching our expectation that larger torques are required to drive larger loads. However, the interesting aspect of this load line is that for significantly large loads, i.e. $D_0 \gg D$, the load line saturates to a limiting slope and the torque equation becomes $\tau = D\Omega_m$, which we refer to as the limiting load-line. The implication of the existence of a limiting load-line is that, if the slope D is less than a critical value $D_c = \tau_{max}/\Omega_m^{(c)}$ then the load line will never intersect the flat region of the TS, i.e. the maximum available torque will not be accessible to the bacterium. We can represent this constraint as:

$$D \geq D_c = \tau_{max}/\Omega_m^{(c)} \qquad (6)$$

Here, D is the rotational drag on flagella and $D_c$ represents the characteristics of the flagellar motor respectively. Thus, Eq. 6 represents a tuned condition of the flagellar geometry with the TS characteristics to enable the bacterium to utilize the maximum torque produced by the flagellar motor for its propulsion. This criterion is applicable only in case of loading the cell-body, instead, if the cargo is loaded on to flagella, then the swim-speeds are lower than body-loading and therefore not desirable.

Based on the intersection of the load-line (Eq. 4) with the TS curve, we obtain the swimming speed V at a given cargo size $D_0$ in terms of maximum torque $\tau_{max}$ and maximum angular speed $\Omega_m^{(max)}$ generated by the BFM and the propulsion matrix elements as:

$$V_{flat} = \frac{B}{A + A_o} \cdot \frac{\tau_{max}}{D} \text{ when } DD_0/(D + D_0) > D_c \qquad (7)$$

$$\text{and } V_{slope} = \frac{B}{A + A_o} \left(\frac{1}{1 + \frac{D}{D_o} + \frac{kD}{D_c}}\right) \Omega_m^{(max)} \text{ when } DD_0/(D + D_0) < D_c \qquad (8)$$

where, $k = (\Omega_m^{(max)} - \Omega_m^{(c)})/\Omega_m^{(c)}$.

When $DD_0/(D + D_0) = D_c$, these two expressions give the same value. The subscripts "flat" and "slope" refers to whether the load-line intersects the TS at its flat region or the downward sloping region. The flat region is where a bacterium can tap the maximum torque produced by the BFM.

For the analysis, we extract the propulsion matrix and TS characteristics of *E. coli* reported in literature namely, $A = 1.48 \times 10^{-8}$ N.s.m$^{-1}$, $B = 7.9 \times 10^{-16}$ N.s, $D = 7 \times 10^{-22}$ N.s.m, $A_o^{cell-body} = 1.4 \times 10^{-8}$ N.s.m$^{-1}$, $D_0^{cell-body} = 4.2 \times 10^{-21}$ N.s.m, $\Omega_m^{(max)} = 900$ rad.s$^{-1}$, $\Omega_m^{(c)} = 450$ rad.s$^{-1}$, $k = 0.5$ [6,14–16] (see Sec. 2 of the SM). Accordingly, the parameter $D_c = \tau_{max}/\Omega_m^{(c)}$ is $2.8 \times 10^{-21}$ N.s.m$^{-1}$ which is about 4 times that of the flagellar rotational drag parameter D. Therefore, we conclude that wild-type *E. coli* doesn't taps the maximum available torque of $\tau_{max} = 1260$ pN.nm [7,17,18], as it doesn't satisfy the matching condition for accessing maximum torque as described in Eq. 6. There are two ways to achieve the matching condition, starting from the wild-type bacteria situation of $D_c \sim 4D$ for *E.coli*. One could either increase D by changing flagellar geometry or decrease $D_c$ by decreasing the maximum torque $\tau_{max}$ or increasing the knee frequency $\Omega_c^{(m)}$. Inspection of Eq. 7 immediately makes it clear that increasing D is not desirable as it reduces the swim speed. Therefore, the latter option of decreasing $D_c$ is preferred. Here again decreasing $\tau_{max}$ is counter-productive as the swim speed is proportional to the torque (Eqs. 7 and 8). Thus, the favourable option is to increase the knee frequency, effectively increasing the constant torque regime of the TS curve. Though there are no reports yet of wilfully shifting the position of the knee frequency in any bacterial system, the TS curves indeed show a shift in knee frequency with change in temperature while maintaining the maximum torque level [17]. Thus, tuning the operating temperature may be one way to tune the flagellar motor characteristics to achieve optimal cargo transport by bacteria.

The swim speeds estimated from Eqs. 7 and 8 as a function of $D/D_c$ for a range of cargo sizes is shown in Fig. 2. In line with the discussion above, we consider two cases to achieve the matching condition of $D/D_c > 1$. Firstly, one can decrease $D_c$ by increasing the knee-frequency for a fixed $\tau_{max}$ and D as shown in Fig. 2(a) which can result in significant improvement $(3 - 4x)$ in swim speed as compared to the unmatched condition for the same cargo size. We refer to this as the optimal situation for cargo transport by flagellated micro-swimmers. Alternately, one can increase D by using a geometric scaling factor, λ, relative to the wild-type dimensions i.e. all geometric dimensions of the wild-type bacteria are multiplied by λ while maintaining the $\tau_{max}$ and $\Omega_c^{(m)}$ at their wild-type values (Fig. 2(b)). Such a transformation will lead to change in A, B and D by factors of $\lambda$ $\lambda^2$ and $\lambda^3$ respectively. There is an optimum geometric scale factor which achieves the maximum swim speed for a given load given by $(D_c/D)^{1/3} = 4^{1/3}$, which is around 1.6 as seen in Fig. 2(b). This optimum scaling factor appears similar to the one found by Purcell [19] but arises out of a completely different reason, namely matching flagellar geometry with the torque-speed curve. Interestingly, the swim speed at the $D = D_c$ condition is around 20 $\mu m.s^{-1}$ close to the observed value for wild-type *E. coli* and *P. aeruginosa*.

The matching condition mentioned in Eq. 6 depends only on the flagellar rotational drag parameter 'D'. One may then assume that the loading of cargo on flagella can circumvent this constraint and achieve higher speeds. However, the speed of bacteria is inversely proportional to the rotational and translational drag on flagella A, D as opposed to only translational drag on bacterial body $A_o$ (Eqs. 7 and 8). Also, the translational drag scales with 'r' whereas rotational drag scales with '$r^3$', hence, if the cargo is loaded on flagella the drag rapidly increases with cargo size resulting in a quick drop of speed as shown in Fig. 2(c). Consequently, the performance of cargo loading on flagella is always worse than cargo loading on bacterial body, i.e. for any given cargo size, body loading achieves higher swim speed than flagellar loading (Fig. 2(d)).

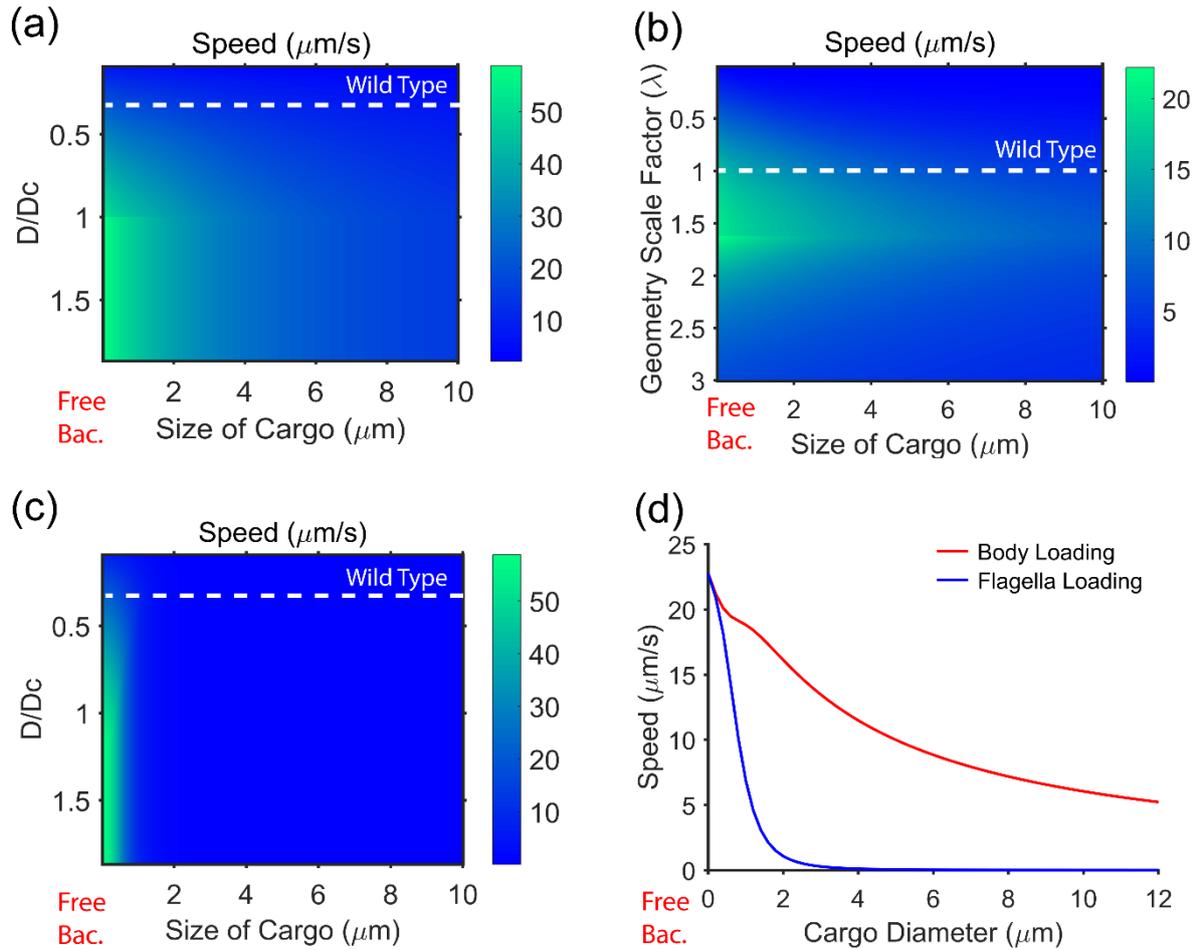

FIG. 2. Surface plot displaying speed of cargo laden bacteria as a function of cargo size. (a) The cargo loaded bacteria achieves the maximum speed at $D = D_c$ and remains constant thereafter. (b) The geometrical parameters A, B and D are scaled as a function of length scale $\lambda$ as ($A = A\lambda$, $B = B\lambda^2$ and $D = D\lambda^3$). Maximum speed is achieved at optimum scaling of geometrical parameters ~1.6 in this case. (c) Flagellar loading is highly inefficient and speed drops rapidly as the size of cargo increases. (d) A comparison of speed vs. cargo loaded on bacterial body and flagella. Evidently speed in the case of flagellar loading drops rapidly.

We examined the validity of our model using experimental measurement of swimming speeds of single bacteria attached with cargos of varying size. Systems with a single bacterium pulling a load are essential to understand general limitations of bio-hybrid devices as they are devoid of complex interactions present in systems with multiple bacteria pulling a single cargo [12]. We used a physical attachment technique for attaching cargo in the form of single oil droplets with diameters up to 12 μm to the head of a single bacterium (see Sec. 3 of the SM). We have used the bacterial species *Pseudomonas aeruginosa* as the model bacterium in this study, which contains $1 - 2$ flagella at one end of its hemi-spherical cap.

The microscopic imaging of freely swimming cargo laden bacteria (Fig. 3(a), SI Video 1) using GFP labelled bacteria with concurrent DIC and fluorescence mode [20] imaging (Fig. 3(b)), revealed the attachment of a single oil drop to a single bacterium with part of the cell body embedded within the oil droplet. The swimming speed of bacteria depends upon the torque and the propulsion force generated by flagellar motor whereas, the swimming efficiency relies on the geometrical parameters (A, B, D) of flagella. These parameters need to be experimentally determined and are not quantified for the bacterial strain *Pseudomonas aeruginosa* (PA14). However, bacterial species PA14 and *E. coli* have similar speed ($\sim 22 \, \mu m.s^{-1}$), body size (l = 1.5 μm, dia. = 0.6 μm) and flagellar length ($\sim 4 \, \mu m$) [14,21]. Hence, as a first approximation we used the values of $A_o^{cell-body}$, $D_o^{cell-body}$, A, B, D from *E. coli* measurements [14]. The TS characteristics is also assumed to be similar to *E. coli* which we take as $\tau_{max} = 1260 \, pN.nm$ [17,18] and $k = 0.5$ [22,23]. The speed vs. cargo size data collected from 33 cargo-laden bacterial trajectories is shown in Fig. 3(c). As explained previously, the speed of cargo loaded bacteria can be estimated using Eq. 8 since, it operates in the slope regime of torque-speed curve under body loaded conditions. We also introduced a correction term to account for increased drag on the bacteria due to their proximity to the glass coverslip as [24]:

$$A_o = A_\infty \left(\frac{1 + 9/16(A_\infty)}{6\pi\eta d_{wall}}\right) \text{ and } D_o = D_\infty \left(\frac{1 + 1/8(D_\infty)}{8\pi\eta d_{wall}^3}\right) \quad (9)$$

where $A_\infty$ and $D_\infty$ are translation and rotational drag coefficients far from the wall, i.e. $A_0^{cell-body}$ and $D_0^{cell-body}$ measured in [14], $\eta$ is the viscosity of surrounding medium and $d_{wall} = 1.5 \, \mu m$ is the 'best fit' average distance of the cargo loaded bacteria from the wall of the fluidic chamber.

The speed estimated using Eq. 8 (red curve) matches closely with experimental data as shown in Fig. 3(c). The incorporation of proximity effect in the form of Eq. 9 resulted in a closer agreement (blue curve) of the model represented by Eq. 8 with experimental measurements. Hence, the measured free-swimming speed and the scaling of swim speed with cargo size is consistent with the picture where the bacteria operate in the regime where only a fraction of the maximum torque from the BFM is utilized. If indeed the bacteria were to operate in the optimal condition, i.e. $D > D_c$, then it would be possible for *P. aeruginosa* (and *E. coli*) to achieve free-swimming speeds of 70-80 $\mu m.s^{-1}$ which is more than three times the experimentally observed value of $\sim 22 \, \mu m.s^{-1}$ (Fig. 3(d)). The variation of swim speed as a function of TS characteristics such as $\tau_{max}$ and $\Omega_c^{(m)}$ are presented in Sec. 4 of the SM.

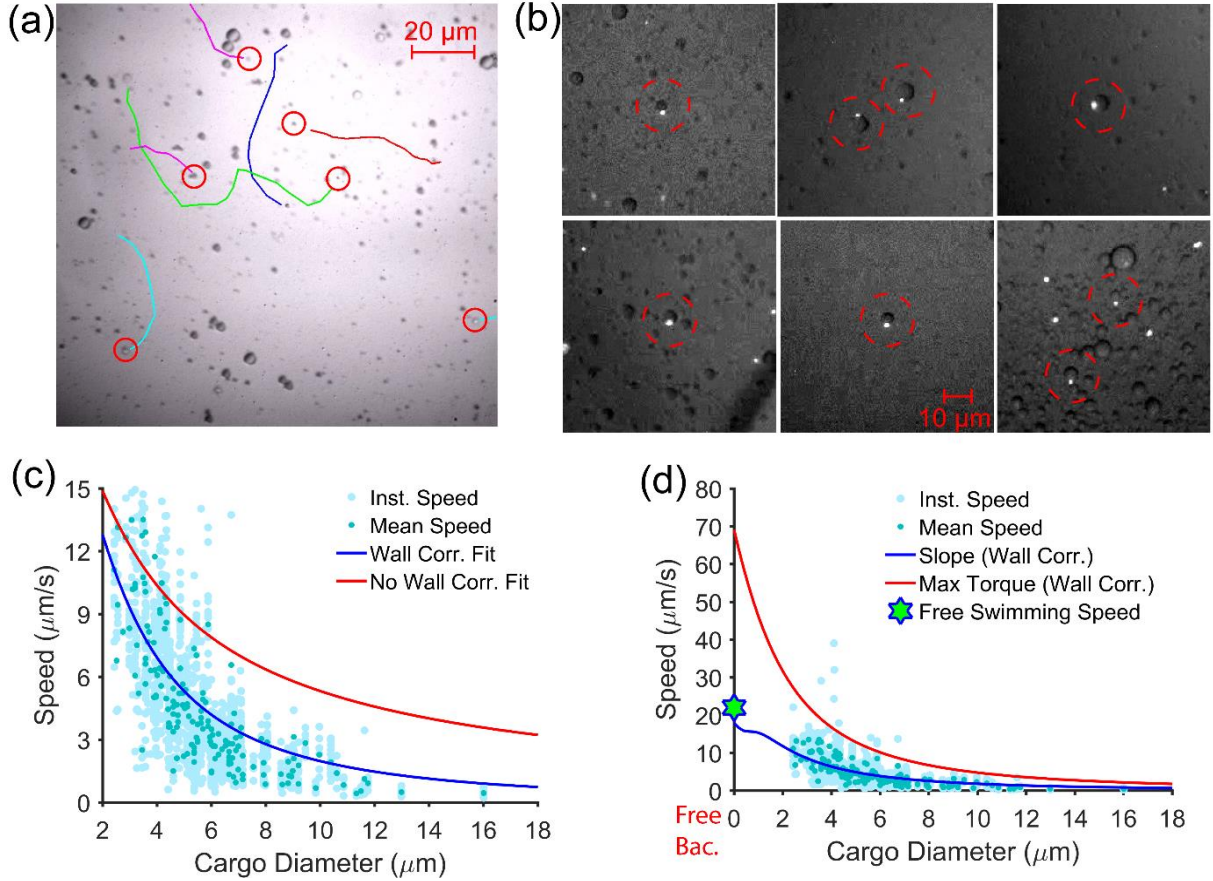

FIG. 3. Swimming dynamics of cargo loaded bacteria. (a) Trajectories of cargo loaded bacteria. (b) Fluorescent mode imaging of bacteria shows bacteria is embedded in the oil droplet. (c) Plot showing average swimming speed vs. cargo size. (d) Graph of speed vs. cargo diameter under maximum torque and slope regime.

Further, the derived expression in Eq. 8 not only gives a general relation to estimate bacterial speed but on a closer inspection also reveals a counterintuitive regime where, swimming speed increases with increasing load. The swimming speed is expected to drop with increase in cargo size whether the cargo is body-loaded or flagellar-loaded, as illustrated in Fig. 2. Instead, we observe anomalous cargo transport regime for $D < D_c$ which can be mathematically understood by re-writing $V_{slope}$ (Eq. 8) under the condition $D/D_0 \gg kD/D_C \gg 1$, which yields:

$$V_{slope} = \frac{B}{A + A_o}\left(\frac{D_o}{D}\right)\Omega_m^{(max)} \qquad (10)$$

The $D_0$ term in the numerator of Eq. 10 scales as $r_0^3$ ($r_0$ is the size of the cargo) while, the $A_0$ term in the denominator scales as $r_0$ leading to the anomalous scaling of swim speed in proportion to $r_0^2$. The condition for the anomalous propulsion regime $\frac{D}{D_o} \gg \frac{kD}{D_c} \gg 1$, requires the existence of an appropriate balance between $k = (\Omega_m^{(max)} - \Omega_m^{(c)})/\Omega_m^{(c)}$ as well as rotational drag on bacterial flagella D and body $D_0$ respectively. Evidently, both k and D should be relatively high to satisfy anomalous propulsion condition requiring low knee frequency $\Omega_m^{(c)}$ and large rotational drag on flagella D. A larger k corresponds to a lower knee frequency $\Omega_m^{(c)}$ resulting in a lower slope of the torque speed curve (Fig. 1(b)). As discussed before, the rotational drag can be increased by scaling up the geometric dimensions by a factor of λ, i.e. $D \rightarrow \lambda^3 D^{(wild-type)}$.

We plot the speed vs. cargo size for low knee frequency $\Omega_m^{(c)} = 100$ Hz and at $\lambda = 1, 3$ in Fig. 4(a) by fixing the maximum BFM torque, $\tau_{max} = 1260$ pN. nm and maximum motor angular speed $\Omega_m^{(max)} = 900$ rad. s$^{-1}$. The anomalous regime where the speed increases with increase in cargo size is not observed at $\lambda = 1$ (wild-type geometry) and is only seen for larger values such as $\lambda = 3$. This is because, as we upscale the size of bacteria by 3 times $\lambda = 3$, the rotational drag coefficient of flagella 'D' grows faster than the rotational drag of bacterial body '$D_0$' eventually satisfying the condition $\frac{D}{D_0} \gg \frac{kD}{D_c} \gg 1$ for anomalous propulsion regime.

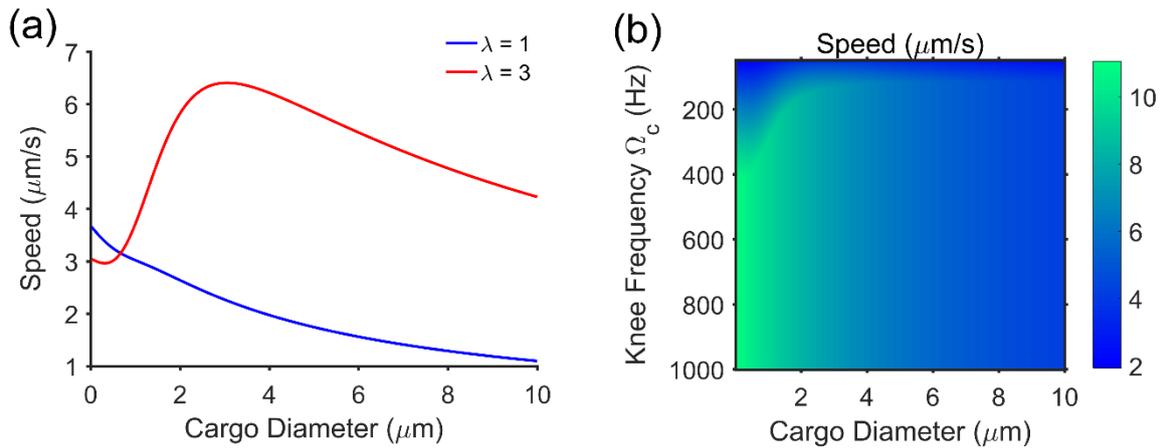

FIG. 4. Anomalous cargo propulsion regime. (a) At a knee frequency of 100 Hz, the geometrical parameters A, B, D are varied using size scaling factor λ. A switch from anomalous propulsion regime to normal behaviour is observed as the λ is increased from 1 to 3. (b) Speed of cargo loaded bacteria increases as the size of cargo is increased for low knee frequency (below 400 Hz).

The plot of speed vs. cargo size at $\lambda = 3$ by varying the knee frequency $\Omega_m^{(c)}$; and as before fixing $\tau_{max} = 1260$ pN. nm, $\Omega_m^{(max)} = 900$ rad. s$^{-1}$ is shown in Fig. 4(b). The anomalous propulsion is observed at lower knee frequency $\Omega_m^{(c)}$ i.e. below 400 Hz. The anomalous regime holds greater significance to the bacterial motion in polymer solutions as it is well known that contrary to the expectation, the swimming speed increases in the presence of polymer solution [7,25]. It is possible for the flagellar bundle to interact sterically in the polymer solutions [26] resulting in a larger rotational drag coefficient of flagella 'D' than the bacterial body '$D_0$' thereby satisfying the condition for anomalous propulsion.

To summarize, we showed the existence of a tuning condition, represented by Eq. 6 which is necessary for bacteria to tap the maximum available torque produced by the flagellar motor. The wild-type bacteria (in particular, *E. coli* and PA14) does not swim under this tuned condition, implying that they are not optimally suited for pulling cargos attached to the cell body. The tuning condition if exploited can lead to an increase by $3 - 4x$ in the swimming speeds relative to the un-tuned wild-type state. This condition $D > D_c$ is independent of the condition required for maximizing propulsion efficiency, namely $A = A_0$. Therefore, speed enhancements and optimal propulsion efficiency can be simultaneously obtained. Our model is consistent with experimental measurements of swimming speed of cargo loaded single bacteria prepared using a novel physical attachment technique. The new expression derived for the speed in slope regime is general and, therefore, quite powerful in predicting previously unknown swimming behaviour such as the anomalous cargo transport regime. These findings are of crucial importance in the design of future bio-hybrid or synthetic flagella-based propulsion systems.

# Supplemental Information

## 1. Expression for swim-speed of cargo loaded bacteria

Force and torque on bacterial body and flagella is given by Purcell's formulation [1]:

$$\begin{pmatrix} F \\ \tau \end{pmatrix} = \begin{pmatrix} A_o & 0 \\ 0 & D_o \end{pmatrix} \begin{pmatrix} V \\ \Omega \end{pmatrix} - \text{body} \quad (S1)$$

$$\begin{pmatrix} -F \\ -\tau \end{pmatrix} = \begin{pmatrix} A & B \\ B & D \end{pmatrix} \begin{pmatrix} V \\ -\omega \end{pmatrix} - \text{flagella} \quad (S2)$$

where $A_o, D_o$ are the translation and rotational drag coefficients and $V, \Omega$ are the translation and angular speed of bacterial body. Similarly, $A, D$ are the translational and rotational drag coefficients and $V, \omega$ the translation and angular speed of flagella. The constant B couples the rotational motion of flagella to the translation motion of bacteria. Force and torque on bacterial body and flagella are equal and opposite since there is no external force. The expression of velocity, force and torque can be written in terms of motor angular speed $\Omega_m = \omega + \Omega$:

$$V = \left[\frac{BD_o}{(A + A_o)(D + D_o) - B^2}\right] \Omega_m \quad (S3)$$

$$\Omega = \left[\frac{(A + A_o)D - B^2}{(A + A_o)(D + D_o) - B^2}\right] \Omega_m \quad (S4)$$

$$\omega = \left[\frac{(A + A_o)D_o}{(A + A_o)(D + D_o) - B^2}\right] \Omega_m \quad (S5)$$

$$F = \left[\frac{A_o B D_o}{(A + A_o)(D + D_o) - B^2}\right] \Omega_m \quad (S6)$$

$$\tau = \left[\frac{(A + A_o)DD_o}{(A + A_o)(D + D_o) - B^2}\right] \Omega_m \quad (S7)$$

For bacteria attached to oil-droplet such as the one we have shown, $D_0 = D_0^{(body)} + D_0^{(cargo)}$ becomes large and hence, the approximation $(A + A_o)(D + D_o) \gg B^2$ is reasonable. A weaker condition $(AD) \gg B^2$ is also satisfied for the experimentally extracted values for *E. coli* $A = 1.48 \times 10^{-8}$ N.$\frac{s}{m}$, $B = 7.9 \times 10^{-16}$ N.s, $D = 7 \times 10^{-22}$ N.s.m. [2]. Accordingly, $B^2$ is ignored resulting in Eqs. S8-S12:

$$V = \left[\frac{BD_o}{(A + A_o)(D + D_o)}\right] \Omega_m \quad (S8)$$

$$\Omega = \left[\frac{D}{(D + D_o)}\right] \Omega_m \quad (S9)$$

$$\omega = \left[\frac{D_o}{(D + D_o)}\right] \Omega_m \quad (S10)$$

$$F = \left[\left(\frac{A_o}{A + A_o}\right)\left(\frac{D_o}{D + D_o}\right)B\right]\Omega_m \tag{S11}$$

$$\tau = \left[\frac{DD_o}{D + D_o}\right]\Omega_m \tag{S12}$$

Equations S8-S12 together with the load line characteristic of bacterial motor can be used to predict the dynamics of cargo carrying bacteria. A typical load-line characteristic of bacterial motor is depicted in Fig. S1.

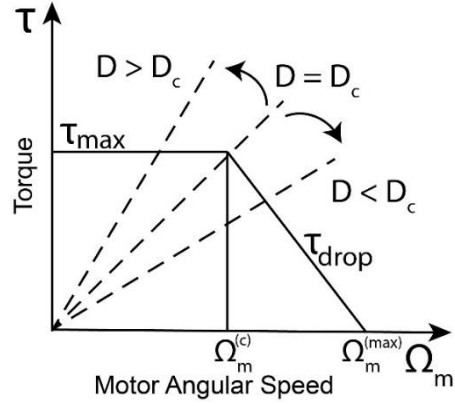

**Figure S1**: Load line characteristic of bacterial motor, where, the motor delivers maximum torque if $D > D_c$.

The speed V (Eq. 8) can be rewritten in terms of torque as:

$$V = \left[\frac{B}{(A + A_o)}\right]\frac{\tau}{D} \tag{S13}$$

If the slope of load line $D > D_c$, then $\tau = \tau_{max}$, hence, the expression for speed is:

$$V_{flat} = \left[\frac{B}{(A + A_o)}\right]\frac{\tau_{max}}{D} \tag{S14}$$

If the slope of load line $D < D_c$, then the exact value of torque ($\tau$) will depend upon intersection of load line with torque speed characteristics of flagellar motor (Fig. S4). The torque from load line and the torque speed curve can be equated as:

$$\left[\frac{DD_o}{D+D_o}\right]\Omega_m = \tau_{max}\left(\frac{\Omega_m^{max} - \Omega_m}{\Omega_m^{max} - \Omega_m^c}\right) \tag{S15}$$

The Eq. 15 can be simplified to extract $\Omega_m$ in terms of load line slope $D_c = \tau_{max}/\Omega_m^c$ and $k = (\Omega_m^{max} - \Omega_m^c)/\Omega_m^c$:

$$\Omega_m = \left(\frac{D_c/k}{\frac{D_c}{k} + \frac{DD_o}{D+D_o}}\right) \tag{S16}$$

The expression of torque (Eq. 12) can be rewritten in terms of $\Omega_m$ as:

$$\tau = \left(\frac{DD_oD_c}{D_c(D+D_o)+kDD_o}\right)\Omega_m^{max} \tag{S17}$$

Hence, for $D < D_c$ speed V can be expressed as:

$$V_{slope} = \frac{B}{A+A_o}\left(\frac{1}{1+\frac{D}{D_o}+\frac{kD}{D_c}}\right)\Omega_m^{(max)} \text{ for } D < D_c \tag{S18}$$

The Eq. 14 and 18 completely determine the speed of bacteria in the regime $D > D_c$ and $D < D_c$ respectively.

## 2. Selection of BFM parameters

The swimming speed of bacteria depends upon the torque generated by flagellar motor ($\tau$, $\Omega_{max}$, k) and the propulsion force generated by bacteria (F) which also relies on the geometrical feature of flagella (A, B, D). For the value of maximum torque $\tau_{max} = 1260 \text{ pN} \cdot \text{nm}$, we relied on the measurements by Berry et al. [3,4]. The geometrical parameters of *E. coli* flagella $A = 1.48 \times 10^{-8} \text{ N.s/m}$, $B = 7.9 \times 10^{-16} \text{ N.s}$, $D = 7 \times 10^{-22} \text{ N.s.m}$ and its maximum angular speed ($\Omega_{max} = 900 \text{ rad/s}$) was determined by Wu et al. [5]. The rotational and translational drag on bacterial body is $A_o^{cell-body} = 1.4 \times 10^{-8} \text{ N.s/m}$, $D_o^{cell-body} = 4.2 \times 10^{-21} \text{ N.s.m}$ respectively. Finally, the value of 'k ≈ 0.4 − 0.6' as per many reported torque speed curves [6,7]. The similar speed (∼ 22 μm/s), size (l = 3 μm, dia. = 0.6 μm), flagellar length (∼ 4 μm) of bacterial species PA14 and *E.coli* [8]; gives confidence in using the values of $A_o^{cell-body}, D_o^{cell-body}, A, B, D$ also for *Pseudomonas aeruginosa* which is explicitly reported only for *E.coli*.

## 3. Attachment of oil-droplet (cargo) on bacteria

Silicone oil droplets is loaded on single PA14 by a simple sonication process. A mixture of 200 μl Silicone oil (viscosity '$\eta_{oil} = 100$ cst') and 10 μl of bacteria (∼$10^6$/mL) in aqueous medium ('$\eta_w = 1$ cst') is shaken using a vortex shaker to form water (containing bacteria) in oil emulsion (Fig. S2(a)). The emulsion is then sonicated in a water bath at a frequency of 40 kHz (BRANSON 2800) for 30 − 45 sec, which facilitates the attachment of oil droplets onto bacteria [9]. Thereafter, water in oil emulsion is centrifuged at 2000 rpm for 1 min resulting in water droplet to settle at the bottom as they are denser (1 gm/ml) than Silicone oil (0.96 gm/ml). The settled water droplet at the bottom consist of bacteria, oil droplets and small oil droplets (∼ 1 μm) attached to the bacteria. The droplets attached to the bacteria act as a seed where other oil droplets condense resulting in loaded bacteria of various possible configurations in about 8 − 9 hrs. The oil droplet loaded bacteria is realised in dilute PBS solution (0.5 mM PBS + 50 μM $MgCl_2.6H_2O$ + 10 μM EDTA) [10] instead of commonly used Berg's motility medium [11]. This is because, at high salt concentrations, these emulsions are unstable resulting in phase separation due to rapid coalescence of oil droplet [12]. The physical attachment of cargo demonstrated here is fundamentally different from chemical attachment techniques demonstrated in previous reports, requiring appropriate viscosity and buffer concentration to ensure the stability of loaded bacteria.

The oil-droplet loaded bacteria is sandwiched between two microscope glass slides forming a microfluidic chamber with height of 50 μm for imaging of bacterial motion. The images of oil droplet loaded bacteria were captured using an Olympus upright microscope (BX51M) simultaneously in DIC and fluorescent modes. A monochrome camera (Olympus – XM10) used with a 20x objective provided a spatial magnification of 0.51 μm per pixel in the imaging plane. The swimming speed of fresh bacteria as estimated from the video analysis of bacterial tracks is 22.9 μm/s which is of the similar order as that of more extensively studied *E. coli* [13]. The mean swim speed of bacteria drops to 15.3 μm/s due to nutrient deprivation during oil droplet condensation phase.

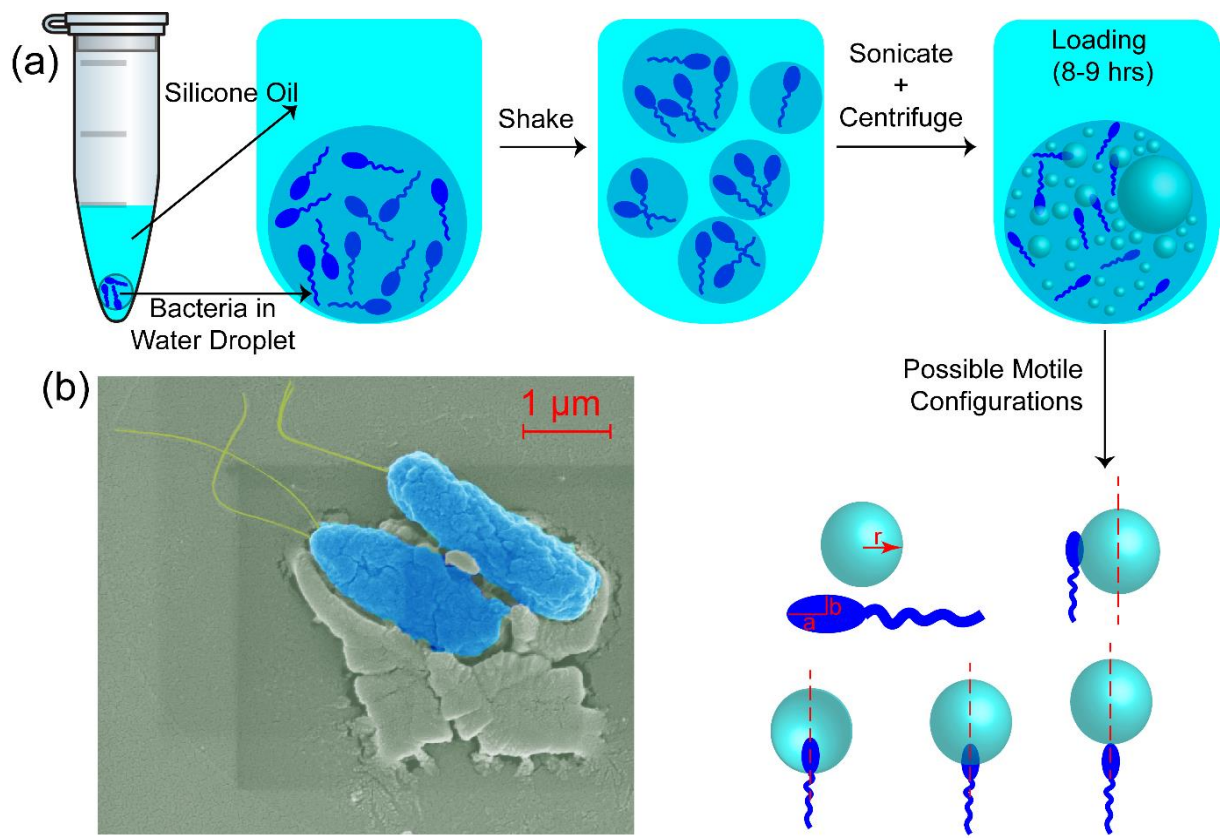

**Figure S2:** Schematic depicting oil-droplet loading process. (a) Initially, 'water containing bacteria' in oil bath is shaken to generate water in oil emulsion which is then sonicated leading to tethering of oil droplet onto bacteria. Over time other smaller oil droplet coalesce on to oil-droplet tethered to bacteria to generate larger loads. (b) SEM image of bacteria reveals PA14 is a biflagellate strain having average length of 1.5 μm and diameter 0.6 μm respectively.

## 4. Effect of parameters on variation of swim-speed with cargo size

The speed of a wild-type swimming bacteria is largely immune with the change in various parameters such as geometry scale factor ($\lambda = 1$), maximum torque ($\tau_{max} = 1260$ pN.nm) and knee frequency ($\Omega_m^c = 600$ rad/s) as shown in Fig. S3(a), S3(b) and S3(c) respectively. The geometry scaling is done as $A_o, D_o, A, B, D \rightarrow A_o\lambda, D_o\lambda^3, A\lambda, B\lambda^2, D\lambda^3$ by keeping all the other parameters same as used in the main text to model Speed vs. Cargo Diameter plot shown in Fig. S3. The knee frequency and maximum torque too is varied similarly by keeping other parameters fixed.

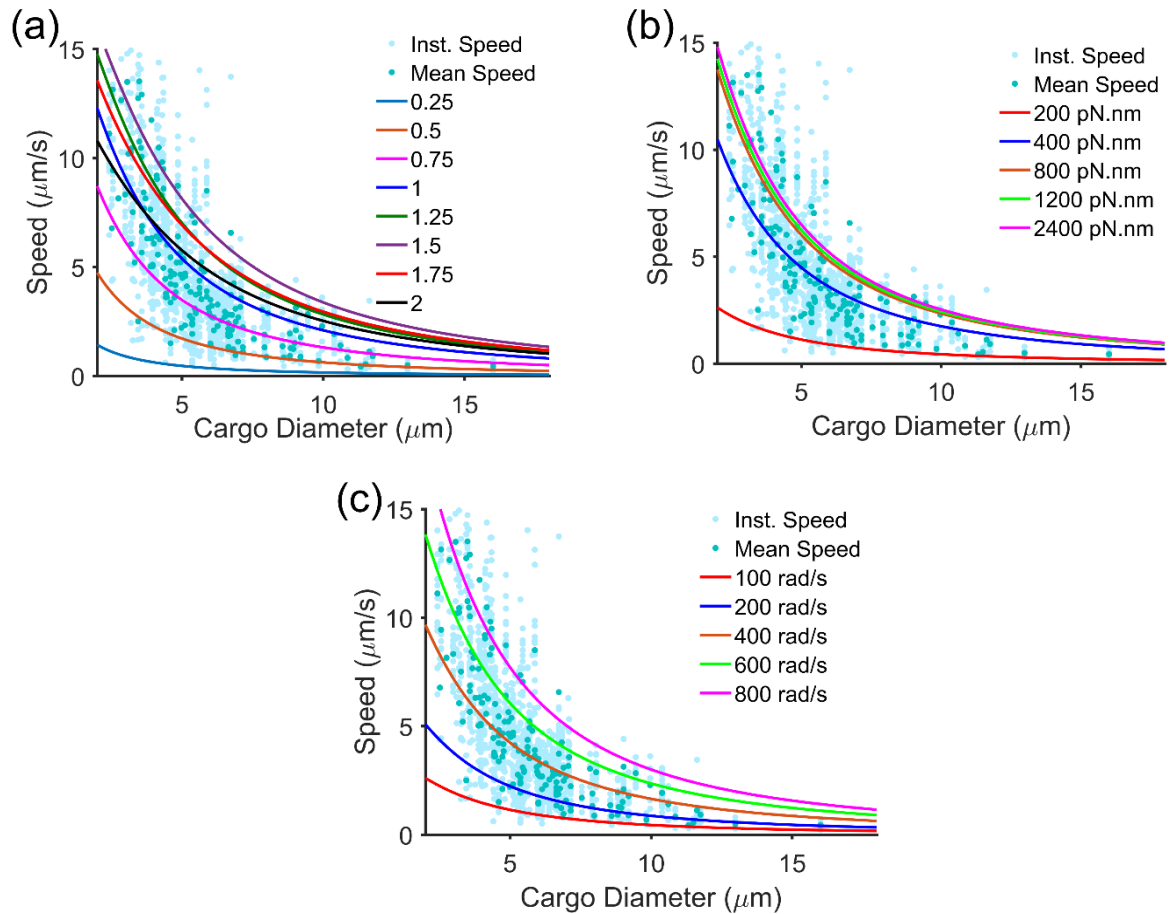

**Figure S3:** Speed as a function of cargo diameter at various parameters. (a) The speed attains maxima at $\lambda = 1.5$ and then drops. (b) Increase in speed saturates around a maximum torque of 800 pN. nm. (c) Speed of bacteria is of the similar order and explains the experimental data for motor speed $400 - 800$ rad/s.

## 5. Description of the SI Videos

SI Video 1: Manually tracked video (DIC mode) of cargo loaded bacteria (frame rate = 4x). Initially, multiple trajectories of cargo loaded bacteria is shown. At 20 sec single trajectory of cargo loaded bacteria shown.